\documentclass[8.5pt,twoside,twocolumn]{article}
\usepackage[super,sort&compress,comma]{natbib} 
\usepackage[version=3]{mhchem}
\usepackage{times}
\usepackage{sectsty}
\usepackage{balance} 
\usepackage{graphicx} 
\usepackage{lastpage}
\usepackage[format=plain,justification=raggedright,singlelinecheck=false,font=small,labelfont=bf,labelsep=space]{caption} 
\usepackage{fancyhdr}
\usepackage{amssymb, amsmath}
\begin{document}
\thispagestyle{plain}
\fancypagestyle{plain}{
\renewcommand{\headrulewidth}{1pt}}
\renewcommand{\thefootnote}{\fnsymbol{footnote}}
\renewcommand\footnoterule{\vspace*{1pt}%
\hrule width 3.4in height 0.4pt \vspace*{5pt}} 
\setcounter{secnumdepth}{5}
\makeatletter 
\def\subsubsection{\@startsection{subsubsection}{3}{10pt}{-1.25ex plus -1ex minus -.1ex}{0ex plus 0ex}{\normalsize\bf}} 
\def\paragraph{\@startsection{paragraph}{4}{10pt}{-1.25ex plus -1ex minus -.1ex}{0ex plus 0ex}{\normalsize\textit}} 
\renewcommand\@biblabel[1]{#1}            
\renewcommand\@makefntext[1]%
{\noindent\makebox[0pt][r]{\@thefnmark\,}#1}
\makeatother 
\renewcommand{\figurename}{\small{Fig.}~}
\sectionfont{\large}
\subsectionfont{\normalsize} 
\fancyfoot{}
\fancyfoot[RO]{\footnotesize{\sffamily{\thepage}}}
\fancyfoot[LE]{\footnotesize{\sffamily{\thepage}}}
\fancyhead{}
\renewcommand{\headrulewidth}{1pt} 
\renewcommand{\footrulewidth}{1pt}
\setlength{\arrayrulewidth}{1pt}
\setlength{\columnsep}{6.5mm}
\setlength\bibsep{1pt}
\twocolumn[
  \begin{@twocolumnfalse}
\noindent\LARGE{\textbf{Quasi-2D dynamic jamming in cornstarch suspensions: visualization and force measurements}}
\vspace{0.6cm}

\noindent\large{\textbf{Ivo R. Peters$^{\ast}$ and
Heinrich M. Jaeger$^{\ast}$}\vspace{0.5cm}}
\vspace{0.6cm}

\noindent \normalsize{We report experiments investigating jamming fronts in a floating layer of cornstarch suspension. The suspension has a packing fraction close to jamming, which dynamically turns into a solid when impacted at a high speed. We show that the front propagates in both axial and transverse direction from the point of impact, with a constant ratio between the two directions of propagation of approximately 2. Inside the jammed solid, we observe an additional compression, which results from the increasing stress as the solid grows. During the initial growth of the jammed solid, we measure a force response that can be completely accounted for by added mass. Only once the jamming front reaches a boundary, the added mass cannot account for the measured force anymore. We do not, however, immediately see a strong force response as we would expect when compressing a jammed packing. Instead, we observe a delay in the force response on the pusher, which corresponds to the time it takes for the system to develop a close to uniform velocity gradient that spans the complete system.}
\vspace{0.5cm}
 \end{@twocolumnfalse}
  ]

\footnotetext{\textit{James Franck Institute \& Department of Physics, The University of Chicago, 929 East 57th Street, Chicago, IL 60637, USA.}} 
\footnotetext{\textit{E-mail: irpeters@uchicago.edu; h-jaeger@uchicago.edu}}

\section{\label{sec:intro}Introduction}
The behavior of water-cornstarch suspensions is often used as a prototypical example of a strong shear thickening fluid.\cite{Fall2008, Brown2009, BischoffWhite2009, Brown2010a, Brown2012a, Fall2012} Of special interest is the limit of discontinuous shear thickening (DST),\cite{Hoffman1972, Fall2008, Brown2009, BischoffWhite2009, Brown2010a, Brown2012a, Fall2012, Fernandez2013, Seto2013, Brown2014} where the stress jumps several orders of magnitude, effectively switching the material from liquid-like to solid-like. A phase diagram of shear thinning, shear thickening, and its relation to jamming was provided by Brown and Jaeger.\cite{Brown2014}

More recently, several experiments have focused on the response of cornstarch suspensions under  normal compression:\cite{Brown2014} \citet{Liu2010} looked at the imprint a sphere makes on a layer of molding clay when it approaches the clay, all submerged in a cornstarch suspension. A surprisingly focused depression was found, unlike what would have been expected for the case of a viscous fluid or dry granular material. \citet{VonKann2011}  showed the appearance of undamped velocity oscillations of a sphere sinking in a bath of cornstarch suspension and stop-go cycles when the sphere approaches the bottom of the container. Shear thickening models were unable to account for their observations, which led to the idea that their system dynamically jammed and unjammed. \citet{Waitukaitis2012} measured the force response on a rod impacting on the surface of a dense cornstarch suspension. Here it was pointed out that it is not shear thickening per se that accounts for the strong force response, but rather the formation of a dynamic jamming front. This jamming front also gives rise to a force response before there is interaction with boundaries. \citet{Roche2013} showed that under sufficient load of an impacting rod, the suspension can even fracture. The speed of propagating crack tips allowed them to estimate an effective shear modulus.

These experiments were all done in a three-dimensional (3D) system. The suspension being opaque, it blocks any visible access to the bulk, which is where the jamming front is propagating. All measurements have been indirect (imprint in clay,\cite{Liu2010} marked wire connected to a settling sphere,\cite{VonKann2011} embedded force sensors\cite{Waitukaitis2012}), or with relatively low temporal and spatial resolution (x-ray at 30 frames/sec).\cite{Waitukaitis2012} Therefore, little is known about the (transient) geometry of the jamming front and what happens with the jammed region when there is interaction with a boundary. 

Using a simple, dry model system of disks, \citet{Waitukaitis2013} provided a more detailed picture for the basic mechanism of how dynamic jamming fronts can develop. Their 2D system starts in an unjammed configuration, which is then uniaxially compressed, giving rise to a traveling front with a speed which is proportional to the pushing speed and diverges as the packing fraction of the initial, undisturbed state approaches jamming. In another 2D system, \citet{Burton2013} showed that jamming fronts play an important role in the dissipation of energy during the collision of dense granular gas clusters.

In this paper, we return to the complexity of a dense cornstarch suspension, but with a setup more in the spirit of Waitukaitis \emph{et al.},\cite{Waitukaitis2013} i.e., we perform impact experiments using a horizontal floating layer of suspension of thickness $h$. This quasi two-dimensional geometry enables us to directly visualize the motion of the suspension using high speed imaging. We then compare our optical measurements to the force we measure as the jamming front is propagating and when it hits a boundary.

The paper is organized as follows: We describe the experimental setup in Sec.~\ref{sec:setup}. In Sec.~\ref{sec:results} we detail the experimental results, which we split up in two parts: the growth of the jammed region before interaction with boundaries (Sec.~\ref{sec:growthRate}-\ref{sec:addedMass}) and after interaction with the boundaries (Sec.~\ref{sec:boundaryInteraction}). We finish with conclusions and discussion in Sec.~\ref{sec:conclusions}.

\section{\label{sec:setup}Experimental setup}
\begin{figure}[htb]
	\centering
	\includegraphics{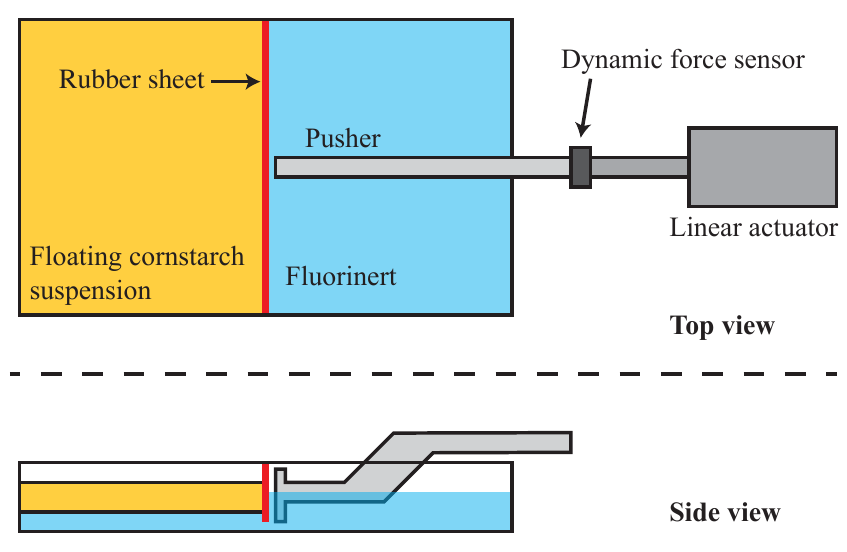}
	\caption{\label{fig:setup}Schematic top view ($a$) and side view ($b$) of the experimental setup. Not shown here are the high speed camera, lights and mirror.}
\end{figure}

Figure~\ref{fig:setup} shows a schematic drawing of the experimental setup. It consists of a rectangular container (30 cm $\times$ 20 cm $\times$ 5 cm) with the bottom filled with fluorinert. Across the container runs a thin rubber sheet (0.13 mm thick silicone rubber, red line in Fig.~\ref{fig:setup}), which restricts the cornstarch suspension to the left half of the container and allows us to perform a normal impact on the suspension from the side.

The layer of suspension has a uniform thickness $h$. We impact the suspension (from right to left in Fig.~\ref{fig:setup}) at controlled speed $u_p$ using a linear actuator (SCN5, Dyadic Systems). The force applied by the linear actuator is measured using a dynamic force sensor (DLC101, Omega). The experiments are recorded with a high speed camera (Phantom V12, Vision Research) at frame rates up to 4 kHz. We obtain a full top-view of the floating suspension with a resolution of $250~\mathrm{\mu m/pixel}$. Tracer particles (ground black pepper) allow us to perform Particle Image Velocimetry (PIV) on the suspension (see for example Fig.~\ref{fig:PIV_example}). 

The suspension is made of cornstarch, water, glycerol, and CsCl. We denote the fraction of cornstarch particles (typical diameter $5$ to $20~\mathrm{\mu m}$ \cite{Fall2008, VonKann2011}) in the total suspension volume by the packing fraction $\phi_0$. We use glycerol to increase the viscosity of the suspending liquid. This does not change the impact behavior,\cite{Waitukaitis2012} but delays the decay of the jammed state and thus enables us to perform the experiments at lower velocities. We density-match the suspension by dissolving CsCl. The density of the suspension is $\rho_s=1580\pm20~\mathrm{kg/m^3}$.

{For our suspensions, we find a spread in measured quantities while keeping $\phi_0$ constant, which is most likely due to variation of the moisture content in the ``dry'' cornstarch grains. When preparing a suspension, we assume that the cornstarch consists of pure dry grains with a density of $1.59\cdot10^3~\mathrm{kg/m^3}$,\cite{Brown2009} so that moisture content will result in an overestimation of the actual packing fraction. We tested this by making two batches of cornstarch which we kept in a humidity controlled chamber: one for $\sim40$ hours at 0.1\%, and one for $\sim100$ hours at 80\% relative humidity. After preparing a suspension with a nominal packing fraction $\phi_0=0.46$ for both batches, we found the values $k=9.9\pm0.4$ and $k=6.6\pm0.2$ (with $k$ the front/pusher velocity ratio, see Sec.~\ref{sec:axialGrowth}) for the batch that was kept at low humidity and high humidity, respectively.\footnote[1]{This is only to illustrate the influence of moisture content on the actual packing fraction. It is not clear how fast cornstarch will loose moisture when kept in a dry environment. These details are outside the scope of this study.} Given the difficulty in precisely determining the absolute packing fraction, we focus in the following on $k$ as the control parameter rather than $\phi_0$.}

Out of plane motions are small (less than 1 mm vertical displacement, measured using a laser sheet),\footnote[2]{The out of plane displacement can also be estimated by adding the decrease in volume on the right side of the rubber sheet to the jammed region, giving a displacement of $\sim0.5~\mathrm{mm}$, in agreement with the laser sheet measurement} and we therefore {do not take it into account in our calculations of the added mass. We note however that the the out of plane displacement is localized in the front region, where the transition from the unjammed to the jammed state occurs. We estimate the maximum out of plane velocity to reach half the pusher speed as the front passes. Once the suspension is jammed, out of plane velocity is negligible.}

In order to approximate a stress-free boundary condition at the bottom surface of the suspension, we use fluorinert (FC-3283) on which the suspension will float (see App.~\ref{app:Fluorinert}). The fluorinert has a density $\rho=1820~\mathrm{kg/m^3}$ and a kinematic viscosity $\nu=0.75~\mathrm{cSt}$.

\section{\label{sec:results}Experimental results}
\begin{figure*}[htb]
	\centering
	\includegraphics{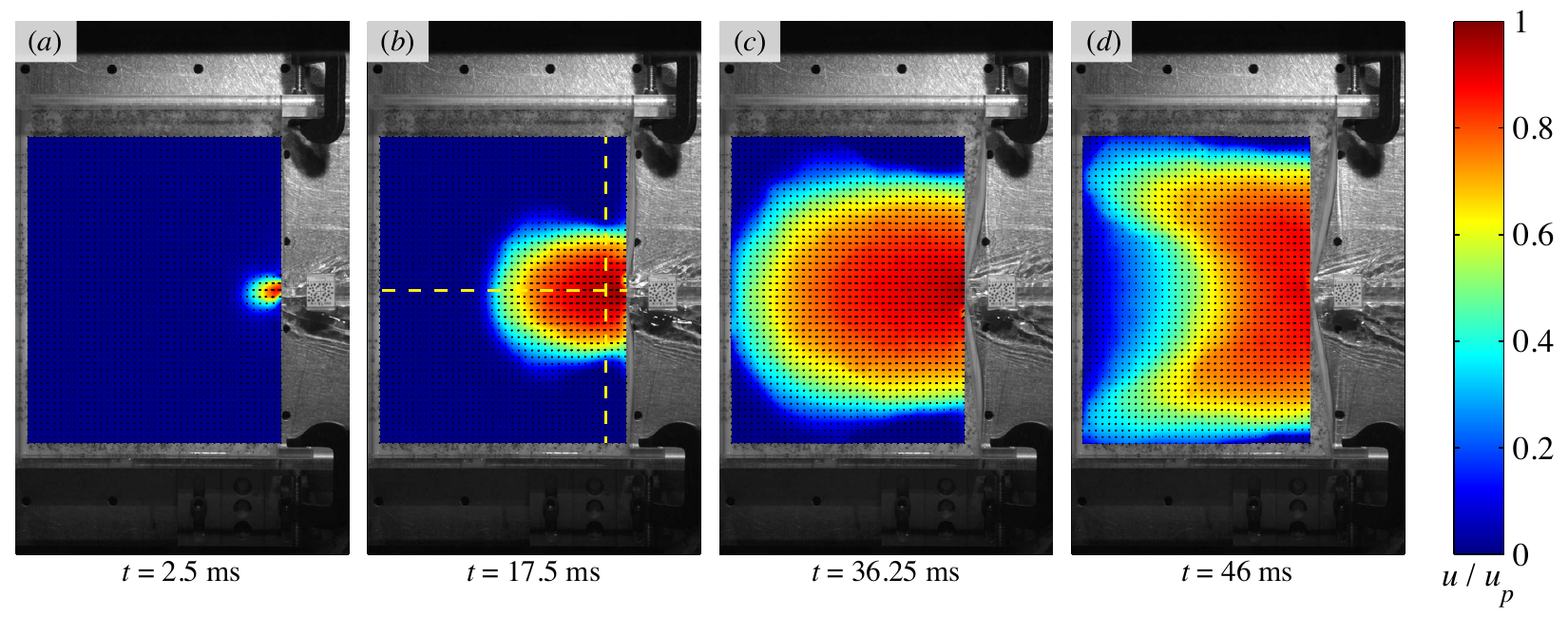}
	\caption{\label{fig:PIV_example}Example of PIV analysis of an experiment. The pusher is moving from right to left with a velocity $u_p=0.375\pm0.007~\mathrm{m/s}$ in ($a$-$c$) and $u_p=0.33~\mathrm{m/s}$ in ($d$). The pusher hits the interface at $t=0$, after which a jammed region starts growing. The colors indicate the magnitude of the axial velocity component (see velocity scale on the right). The dashed lines in ($b$) indicate the position where the velocity profiles of Fig.~\ref{fig:velocityProfiles} are calculated. See supplementary material for a movie.}
\end{figure*}
In every experiment, the pusher approaches the suspension at a constant speed $u_p$ from the right. When the pusher hits the suspension, the motion initially is localized around the impact site (Fig.~\ref{fig:PIV_example}[$a$]). Fig.~\ref{fig:PIV_example} shows the axial component of the velocity $u$, normalized by the pusher velocity $u_p$. As the pusher moves in further, the motion in the suspension grows in both the axial and transverse direction (Fig.~\ref{fig:PIV_example}[$b$]). Remarkably, the gradient from high to low velocity is very localized, as can be seen in Fig.~\ref{fig:velocityProfiles}. We identify a dynamically jammed region as the part where the suspension approaches uniform motion, i.e., velocity gradients are moderate. We define the dynamic jamming front \cite{Waitukaitis2012,Waitukaitis2013} as the position where the velocity of the suspension is half that of the pusher. The front travels at a speed $u_f$, much larger than the pushing speed $u_p$ (see section \ref{sec:axialGrowth}), thereby expanding the jammed region.

The jammed region keeps growing until the front reaches the solid boundary (Fig.~\ref{fig:PIV_example}[$c$]), after which the system has to adjust to new boundary conditions. Visually from the PIV results, this adjustment looks like a stagnation front that moves from left to right towards the pusher (Fig.~\ref{fig:PIV_example}[$d$]). Our physical interpretation of this changing velocity field can be found in Sec. \ref{sec:boundaryInteraction}.

We will first focus on the growth of the jammed region before there is any interaction with boundaries. The two main points of interest are (i) the growth rate in axial and transverse direction, and (ii) the force response on the pusher. Following Waitukaitis and Jaeger,\cite{Waitukaitis2012} there is a growing force response due to an added mass term. In our quasi-2D system, we can visualize the motion of the suspension, and therefore accurately measure the added mass term by integrating the velocity field. We will shortly discuss the role of dissipation during the growth of the jammed region. After this, we will consider the response of the system when there is interaction with solid boundaries.

\subsection{\label{sec:growthRate}Growth rate}
We first quantify the growth of the jammed region through analysis of the PIV data, starting with the axial growth (Sec.~\ref{sec:axialGrowth}), and then the transverse growth (Sec.~\ref{sec:transverseGrowth}).

\subsubsection{\label{sec:axialGrowth}Axial growth.~~}
\begin{figure}[htb]
	\includegraphics{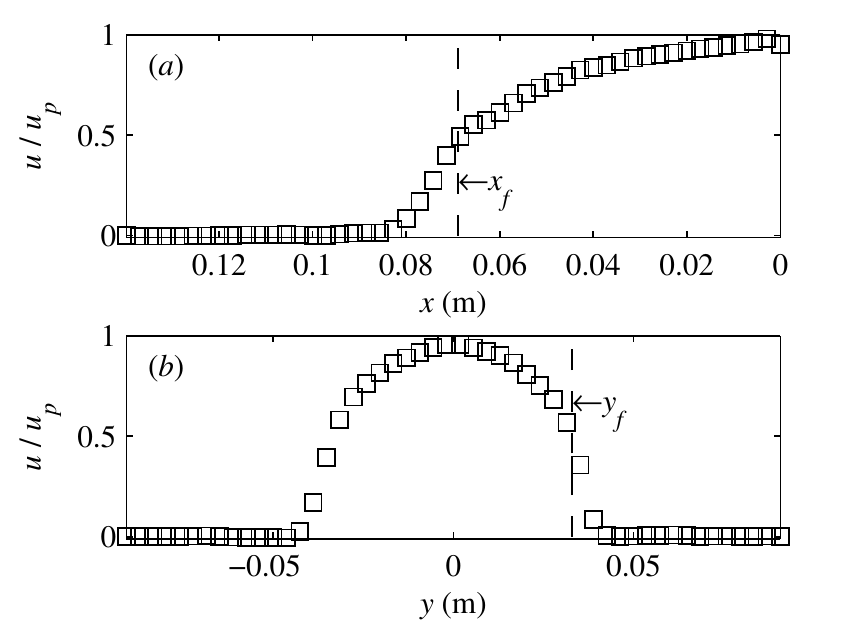}
	\caption{\label{fig:velocityProfiles}Axial ($a$) and transverse ($b$) velocity profiles at $t=17.5~\mathrm{ms}$ along the lines indicated in Fig.~\ref{fig:PIV_example}($b$). The vertical dashed lines indicate the axial and transverse front position. {The maximum slope $du/dx$ of the axial velocity in ($a$) is approximately $-14~\mathrm{s^{-1}}$. At the transverse front position $y_f$, the shear rate (slope of the velocity profile $du/dy$) is approximately $15~\mathrm{s^{-1}}$.}}
\end{figure}

\begin{figure}[htb]
	\includegraphics{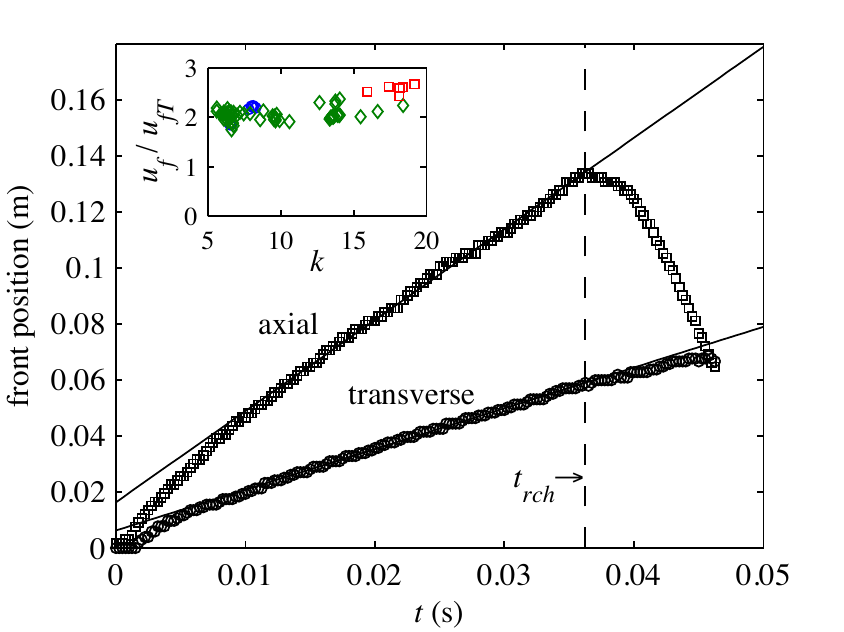}
	\caption{\label{fig:axialTransverseGrowth}Example of the axial and transverse growth of the jammed region. The solid lines are fits to the linear regime. The velocity of the axial front is $2.83~\mathrm{m/s}$; the transverse front velocity is $1.52~\mathrm{m/s}$. The vertical dashed line indicates $t_{rch}$, when the axial front reaches the wall of the container. There is a initial transient time ($t\lesssim0.01$) where the front develops (see main text). Inset: Ratio between axial and transverse front speed as a function of $k$. Blue circles: $\phi_0=0.43$; green diamonds: $\phi_0=0.46$; red squares: $\phi_0=0.48$. }
\end{figure}

We measure the growth in the axial direction with the following 5 steps: (i) We determine the time window during which there is a forward (from right to left in Fig.~\ref{fig:PIV_example}) motion of the jamming front. (ii) By tracking the position of the pusher within this interval we determine $u_p$ and the standard deviation of $u_p$. (iii) Using the PIV-analysis, we determine the normalized velocity profile $u/u_p$ for each frame (see for example Fig.~\ref{fig:velocityProfiles}[$a$]). (iv) Defining the front position by the point where $u/u_p=1/2$,\footnote[3]{The front velocities as measured in Fig.~\ref{fig:axialTransverseGrowth} are insensitive to the choice of cut-off velocity for the determination of the front position.} we determine the position of the front as a function of time (see Fig.~\ref{fig:axialTransverseGrowth}). (v) A fit to the linear regime gives the front velocity $u_f$. In Fig.~\ref{fig:axialTransverseGrowth}, the linear regime is $10~\mathrm{ms}\lesssim t<t_{rch}$, where $t_{rch}$ is the moment when the front reaches the wall of the container (see Sec. \ref{sec:boundaryInteraction}).

In the example of Fig.~\ref{fig:axialTransverseGrowth}, we find a ratio $k$ between the pusher velocity and the front velocity of $k=u_f/u_p-1=6.5$. Using $k=\frac{\phi_0}{\phi_J-\phi_0}$,\cite{Waitukaitis2013} with $\phi_0=0.43$ the initial, as prepared packing fraction and $\phi_J$ the jamming packing fraction, gives $\phi_J=0.50$. Collecting data from all our experiments, we arrive at a value $\phi_J=0.51\pm0.02$, which is consistent with data from 3D-experiments in cornstarch.\cite{Waitukaitis2012,Waitukaitis} 

A closer look at the axial velocity profile in Fig.~\ref{fig:velocityProfiles} shows three distinct regions. Going from left to right, far away from the pusher, at $x\gtrsim0.08~\mathrm{m}$, the velocity of the suspension is uniformly zero. In the center, at $0.05~\mathrm{m}\lesssim x \lesssim 0.08~\mathrm{m}$, there is a region with a large velocity gradient (front region), which connects the fast moving region with the quiescent region. The velocity gradient indicates the compaction which happens when going from an unjammed to a jammed state.\cite{Waitukaitis2013} The width $w$ of the front is of the order of $2~\mathrm{cm}$. The front will need time to develop, which we estimate by using the speed $u_f$ at which the front travels. This gives a typical time $w/u_f\approx7~\mathrm{ms}$, which corresponds to the transient time in Fig.~\ref{fig:axialTransverseGrowth}. Finally, there is a region with a small but significant velocity gradient. 

A possible explanation why we find a velocity gradient in the jammed region is that the stress is not constant in time. All the stress that is applied along the boundary of the jammed region is transferred backwards and accumulates at the pusher. This translates into an additional compaction inside the jammed region. Just like the compaction from unjammed to jammed results in a velocity gradient in the front region, the slight additional compaction in the jammed region results in a small velocity gradient. As we will show in Sec. \ref{sec:boundaryInteraction}, this velocity gradient will increase and span the complete system once the stress increases due to interaction with the opposing boundary.

\subsubsection{\label{sec:transverseGrowth}Transverse growth.~~}
We analyze the transverse growth in the same way as we do for the axial growth, and find that it has similar properties. Fig.~\ref{fig:velocityProfiles}($b$) shows the velocity profile along the vertical dashed line shown in Fig.~\ref{fig:PIV_example}($b$). Note that the velocity plotted in Fig.~\ref{fig:velocityProfiles}($b$) is still the axial component of the velocity (i.e., $u$).

There are two transverse fronts, which move fairly symmetrically sideways from the point of impact at $y=0$. We determine the position of the transverse fronts by taking their average position with respect to the origin. Fig.~\ref{fig:axialTransverseGrowth} shows that the transverse fronts move with a constant velocity $u_{fT}$. The value we find for $u_{fT}$ is insensitive to the exact position where we determine the velocity profiles (vertical dashed yellow line in Fig.~\ref{fig:PIV_example}[$b$]). In Figs.~\ref{fig:velocityProfiles}-\ref{fig:axialTransverseGrowth}, we have determined the velocity profiles at a distance of $7~\mathrm{mm}$ from the rubber sheet. Measuring the velocity profiles at double ($14~\mathrm{mm}$) or half ($3.5~\mathrm{mm}$) the distance from the rubber sheet gives a difference less than 1\% in $u_{fT}$. We find that for our experiments the ratio between the axial and transverse velocity of the jamming front ranges from 1.7 to 2.7. The ratio increases weakly with $k$, see inset Fig.~\ref{fig:axialTransverseGrowth}. We have performed experiments with a rubber sheet of 4 times the thickness of the one used in the rest of the experiments, and found the same ratio ($2.01\pm0.03$ for the experiments performed).

When we examine the transverse velocity profile the same way we did for axial profile, we find also here a region, far away from the pusher, where the velocity is uniformly zero. Then there is a region with a large velocity gradient, followed by a smaller velocity gradient closer to the pusher. Although the velocity profiles show similar features, the interpretation of the velocity gradients in the transverse direction is different from the axial one. In the axial direction, a velocity gradient indicates that there is compression in the front. In the transverse direction however, the velocity gradients are shear flows. The typical magnitude of the shear rate in Fig.~\ref{fig:velocityProfiles}($b$) is $\dot{\gamma}=\frac{\partial u}{\partial y}$ = $15\pm2~\mathrm{s^{-1}}$. 

\subsubsection{General shape.~~}
\begin{figure}[htb]
	\centering
	\includegraphics{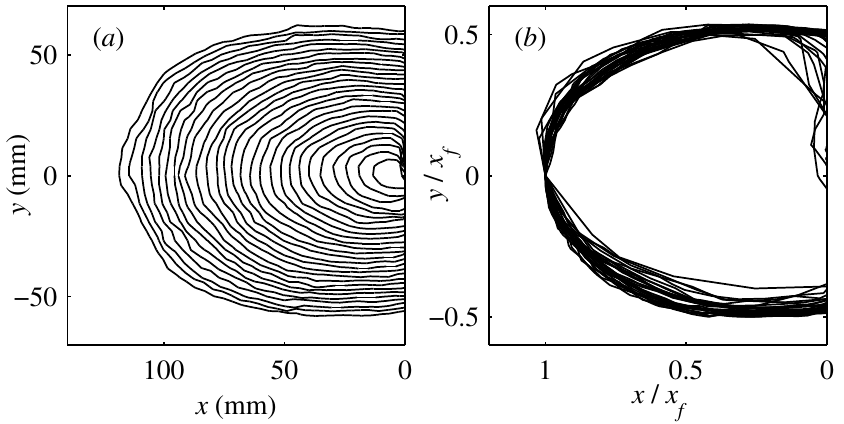}
	\caption{\label{fig:frontShapes}Shape of the front plotted for different instances of time. Time difference between each line is $2.5~\mathrm{ms}$. Lines are drawn where the axial component of the velocity is half the pusher velocity (i.e., $u/u_p=\frac{1}{2}$). The origin of the coordinate system is defined as the instantaneous position of the pusher. ($a$): the evolution of the front during the experiment. ($b$): the same lines, normalized by the axial front position $x_f$.}
\end{figure}
Above, we determined the position of the jamming front in the axial and transverse direction. We can apply the same approach to arbitrary directions, which results in a well-defined shape of the jammed region. Fig.~\ref{fig:frontShapes}($a$) shows the development of the jamming front through contour lines where $u=u_p/2$ for a time interval of about $35~\mathrm{ms}$. In Fig.~\ref{fig:frontShapes}($b$) we normalize the contour lines by the front position $x_f$, demonstrating that the front has essentially a self-similar shape, while the overall size (defined by the length scale $x_f$) grows almost one order of magnitude.

\subsection{\label{sec:addedMass}Added mass}
\begin{figure}[htb]
	\includegraphics{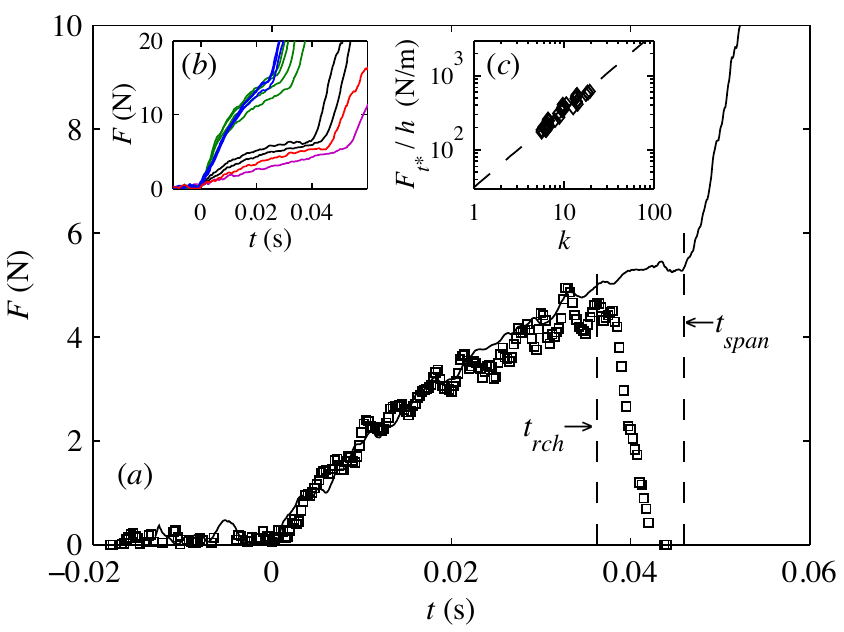}
	\caption{\label{fig:forceComparison}($a$) Direct comparison, without adjustable parameters, of the force calculated from momentum conservation (squares) and the force measured with the force sensor (solid line). Only after interaction with the boundary ($t\approx t_{rch} = 0.036~\mathrm{s}$) do the values deviate. ($b$) Representative set of measured forces as a function of time for different experimental conditions. Black: $h=16.5~\mathrm{mm}$, $k=8.0\pm0.9$, $\nu=6.9~\mathrm{cSt}$; Green: $h=30.0~\mathrm{mm}$, $k=12.1\pm0.8$, $\nu=7.2~\mathrm{cSt}$; Purple: $h=13.0~\mathrm{mm}$, $k=6.0$, $\nu=43~\mathrm{cSt}$; Blue: $h=22.0~\mathrm{mm}$, $k=17.9\pm0.5$, $\nu=7.8~\mathrm{cSt}$; Red: $h=15.0~\mathrm{mm}$, $k=7.5$, $\nu=6.9~\mathrm{cSt}$. ($c$) Force per unit suspension depth at time $t^*=0.02~\mathrm{s}$ as a function of $k$. The dashed line is a fit with slope 1.0. }
\end{figure}

During the experiment we find a growing force response of the suspension on the pusher (Fig.~\ref{fig:forceComparison}($a$,$b$), solid lines). A similar behavior was found by \cite{Waitukaitis2012} in a 3D experiment, through the deceleration of the impacting rod. There are different possible explanations for the origin of this rather strong response. Considering the cornstarch suspension as a strongly shear thickening liquid, viscous interactions with the walls would be a first possibility. A different possibility is that interaction with the walls can be neglected, and the force originates purely from added mass due to the growing jammed region. Figures~\ref{fig:PIV_example}-\ref{fig:velocityProfiles} suggest the latter explanation, because there is no significant velocity gradient near the walls. Viscous interaction with the boundaries would only be possible if there is a velocity gradient which connects to the wall.

We can calculate the contribution of added mass by integrating the velocity fields we obtained through the PIV-analysis. From every frame, we determine the momentum of the system by evaluating $p(t)=\rho_s h \iint u dxdy$, with $\rho_s$ the density of the suspension and $h$ the thickness of the suspension layer. After this, we calculate the force by taking the time derivative $F=\frac{\partial p}{\partial t}$, assuming conservation of momentum within the suspension. In Fig.~\ref{fig:forceComparison}($a$) we compare the force we obtain through momentum conservation with the actual force measured with a dynamic force sensor.

If there would exist an interaction with the wall, either through the suspension, or through the rubber sheet, a part of the force would would not go into momentum of the suspension, but rather to the boundaries. In that case, the momentum balance described above would show a difference between the calculated force and the signal from the force sensor. Clearly, Fig.~\ref{fig:forceComparison}($a$) shows that there is no significant contribution from the boundaries, for $t\lesssim0.35$. Similarly, Fig.~\ref{fig:forceComparison}($a$) shows that no significant amount of momentum is transferred to the fluorinert (see also App. \ref{app:Fluorinert}).

The two-dimensional nature of our experiment suggests that the force curves can be normalized by the thickness $h$ of the suspension layer. Indeed, Fig.~\ref{fig:forceComparison}($c$) shows that the forces measured at a fixed time $t^*$ follow the same trend with $k$ when normalized by $h$. An increase of the measured force is expected for increasing $k$, because the added mass grows faster for higher values of $k$.

\subsection{\label{sec:boundaryInteraction}Interaction with boundaries}
\begin{figure}[htb]
	\includegraphics{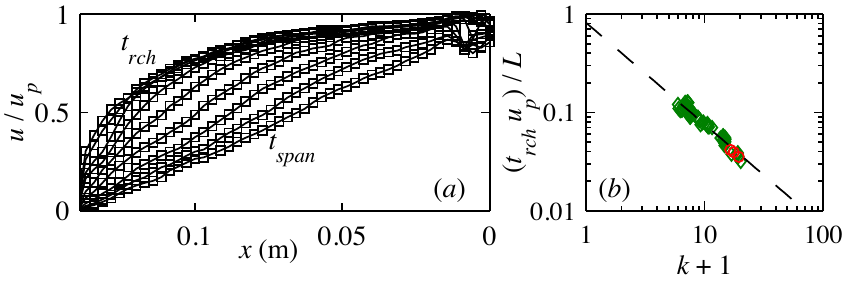}
	\caption{($a$) Evolution of the velocity profiles between $t_{rch}$ (top curve) and $t_{span}$ (bottom curve). Time interval between different profiles is $1.25~\mathrm{ms}$. ($b$) Time to reach system boundary, $t_{rch}$, made dimensionless by the ratio of size of the experiment $L$ to pusher velocity $u_p$, as function of $k+1$. Slope of dashed line is -1.0. Green diamonds: $L=0.150~\mathrm{m}$, red circles: $L=0.077~\mathrm{m}$.}
	\label{fig:boundaryInteraction}
\end{figure}

The results we described above are all during the time that there is no interaction with any of the boundaries. Figure~\ref{fig:PIV_example}($d$) shows that the structure of the velocity field changes drastically after interaction with the boundaries. In Fig.~\ref{fig:axialTransverseGrowth} we defined the time $t_{rch}$ as the moment when the detected front position reaches a maximum. We confirm that this behavior is due to the interaction with the opposite wall, and that the side walls can be neglected, by performing the experiment with half the distance between the rubber sheet and the opposite wall $L$, and keeping the distance between the side walls fixed. In Fig.~\ref{fig:boundaryInteraction}($b$) we show that, as expected, $t_{rch}u_p/L\propto(k+1)^{-1}$. Note that to estimate the time that the front reaches the boundary, we have to use $k+1$: $k$ only represents the growth of the solid region, but the solid region is also translated as a whole at a speed $u_p$, which gives $k+1= u_f/u_p$.

The interaction can be broken up in two steps: First, as the jamming front reaches the boundary, the velocity profile changes, which can be detected as a front traveling back towards the pusher (see the axial front position for $t>t_{rch}$ in Fig.~\ref{fig:axialTransverseGrowth}). After this, once a uniform velocity gradient that spans the system size has been established, we detect a strong response at the force sensor.

Figure~\ref{fig:boundaryInteraction}($a$) shows the evolution of the velocity profile for times larger than $t_{rch}$. The front stops moving from right to left, but instead starts to flatten out. The result of this is that the detected front position ($u/u_p=\frac{1}{2}$) starts to move from left to right. Fig.~\ref{fig:forceComparison}($a$) also shows that for $t>t_{rch}$, there is a large discrepancy between the force calculated through momentum conservation and the force measured by the force sensor, because momentum is being transferred to the walls of the container. However, although there is a clear interaction with the wall, there is no clear sign of this in the \emph{measured} force at $t=t_{rch}$ (see Fig.~\ref{fig:forceComparison}[$a$]).

If we imagine the response of a homogeneous solid between the pusher and the wall, the expected velocity (or displacement) profile would be linear, due to the two boundary conditions $u=u_p$ at $x=0$ and $u=0$ at $x=L$ (with $L$ the distance between the pusher and the wall). Clearly, we observe a very different velocity profile at $t=t_{rch}$.
To quantify the difference of the observed profile with the profile expected for a homogeneous solid for $t>t_{rch}$, we calculate the difference between our velocity profile and a purely linear one, $u/u_p=1-x/L$. We define $t_{span}$ as the moment when the difference (least squares) between the measured profile and the linear profile is minimal. Figure~\ref{fig:forceComparison}($a$) shows that for $t>t_{span}$ there is a sudden large increase in force, much like the behavior expected for a solid connecting the pusher with the wall.

\section{\label{sec:conclusions}Conclusions and discussion}
We have directly visualized dynamic jamming fronts in a dense cornstarch suspension (Fig.~\ref{fig:PIV_example}). The front propagates at a constant speed in axial and transverse direction (Fig.~\ref{fig:axialTransverseGrowth}), resulting in a self-similar shape (Fig.~\ref{fig:frontShapes}). The constant speed with a ratio $k$ depending on how close the system is to the jamming point agrees at least qualitatively with the 2D model system of Waitukaitis \emph{et al.},\cite{Waitukaitis2013} pointing out the difference with shocks that travel through a system beyond the jamming point.\cite{Gomez2012, Gomez2012a} Our results demonstrate that the force response on the impactor can be completely accounted for by the added mass until the front reaches the system boundaries (Fig.~\ref{fig:forceComparison}). From the axial velocity profile we identify three regions, going from (i) unjammed to (ii) jammed, and (iii) further compression of a jammed packing (Fig.~\ref{fig:velocityProfiles}). Finally, we observed that after the moment of first interaction with the boundaries, there is a delay in the response of the force sensor. This delay corresponds to the time it takes for the jammed region to become homogeneous across the full system (Fig.~\ref{fig:boundaryInteraction}).

Our experiments confirm the added mass model proposed earlier,\cite{Waitukaitis2012} but the shape of the jammed solid appears to be rather different. An important remaining question therefore is how the shapes in Fig.~\ref{fig:frontShapes} map to a 3D experiment. Related to that, it is of interest to know whether the ratio between $u_f$ and $u_{fT}$ we observe is a feature of the two-dimensional nature of the current experiment or more general. To answer these questions, a mechanism that explains the transverse growth rate of the jammed region will be needed. We speculate that it might be possible to think of this as dynamic shear jamming, whereby instead of having a fixed volume that is quasi-statically deforming, as in ordinary shear-jamming,\cite{Bi2011, Sarkar2013, Seto2013} the confinement is a dynamic one resulting from the inertia of the quiescent, unjammed suspension. Finally, while the experiments reported here focused on the growth of the impact-generated jammed region and its interaction with system boundaries, they did not explore the eventual fracture of such a jammed solid as the applied stress is increased even further. This has been investigated in 3D systems by analyzing the crack pattern on the impacted surface.\cite{Roche2013} 
However, in the current experimental setup, the required magnitude of force will buckle the floating layer and hence the conditions are not well controlled. An adjusted setup (for example a thicker suspension layer or a smaller system size) possibly omits this problem and would allow for a way to investigate how cracks propagate into a jammed solid.
\balance

\section*{Acknowledgements}
We thank Eric Brown, Endao Han, Victor Lee, Sayantan Majumdar, Marc Miskin, Shomeek Mukhopadhyay, Scott Waitukaitis, and Tom Witten for insightful discussions. We wish to thank the referees for their helpful suggestions to improve the original manuscript. This work was supported by the US Army Research Office through grant W911NF-12-1-0182.

\appendix
\section{\label{app:Fluorinert}Influence of fluorinert}
We estimate an upper bound for the influence of the fluorinert by investigating the added mass and estimate the viscous dissipation. For simplicity, we will treat the floating cornstarch suspension as a moving solid boundary that drives the fluid below it. Momentum is therefore transferred to the fluorinert through the growing boundary layer with thickness $\delta \sim \sqrt{\nu t}$, where $\nu$ is the kinematic viscosity of the fluorinert, and $t$ the typical time of the experiment. Taking $t\approx0.06~\mathrm{s}$ gives $\delta\sim0.2~\mathrm{mm}$. With a layer of cornstarch suspension of $15~\mathrm{mm}$, the moving mass of fluorinert compared to the suspension is $\lesssim2\%$.

Considering the viscous stress due to the fluorinert, using an average boundary layer $\delta\sim0.1~\mathrm{mm}$ and a velocity $U_0=0.4~\mathrm{m/s}$ gives $\tau\sim\rho\nu U_0/\delta\sim5.6~\mathrm{Pa}$, acting on a typical surface of 0.2 m by 0.15 m results in a typical force $F_v\sim0.2~\mathrm{N}$. This upper bound is 5\% of the typical force response measured in the experiment, and falls within the noise of our measurements.

\footnotesize{
\bibliography{} 

\providecommand*{\mcitethebibliography}{\thebibliography}
\csname @ifundefined\endcsname{endmcitethebibliography}
{\let\endmcitethebibliography\endthebibliography}{}
\begin{mcitethebibliography}{21}
\providecommand*{\natexlab}[1]{#1}
\providecommand*{\mciteSetBstSublistMode}[1]{}
\providecommand*{\mciteSetBstMaxWidthForm}[2]{}
\providecommand*{\mciteBstWouldAddEndPuncttrue}
  {\def\EndOfBibitem{\unskip.}}
\providecommand*{\mciteBstWouldAddEndPunctfalse}
  {\let\EndOfBibitem\relax}
\providecommand*{\mciteSetBstMidEndSepPunct}[3]{}
\providecommand*{\mciteSetBstSublistLabelBeginEnd}[3]{}
\providecommand*{\EndOfBibitem}{}
\mciteSetBstSublistMode{f}
\mciteSetBstMaxWidthForm{subitem}
{(\emph{\alph{mcitesubitemcount}})}
\mciteSetBstSublistLabelBeginEnd{\mcitemaxwidthsubitemform\space}
{\relax}{\relax}

\bibitem[Fall \emph{et~al.}(2008)Fall, Huang, Bertrand, Ovarlez, and
  Bonn]{Fall2008}
A.~Fall, N.~Huang, F.~Bertrand, G.~Ovarlez and D.~Bonn, \emph{Phys. Rev.
  Lett.}, 2008, \textbf{100}, 018301\relax
\mciteBstWouldAddEndPuncttrue
\mciteSetBstMidEndSepPunct{\mcitedefaultmidpunct}
{\mcitedefaultendpunct}{\mcitedefaultseppunct}\relax
\EndOfBibitem
\bibitem[Brown and Jaeger(2009)]{Brown2009}
E.~Brown and H.~M. Jaeger, \emph{Phys. Rev. Lett.}, 2009, \textbf{103},
  086001\relax
\mciteBstWouldAddEndPuncttrue
\mciteSetBstMidEndSepPunct{\mcitedefaultmidpunct}
{\mcitedefaultendpunct}{\mcitedefaultseppunct}\relax
\EndOfBibitem
\bibitem[{Bischoff White} \emph{et~al.}(2009){Bischoff White}, Chellamuthu, and
  Rothstein]{BischoffWhite2009}
E.~E. {Bischoff White}, M.~Chellamuthu and J.~P. Rothstein, \emph{Rheol. Acta},
  2009, \textbf{49}, 119--129\relax
\mciteBstWouldAddEndPuncttrue
\mciteSetBstMidEndSepPunct{\mcitedefaultmidpunct}
{\mcitedefaultendpunct}{\mcitedefaultseppunct}\relax
\EndOfBibitem
\bibitem[Brown \emph{et~al.}(2010)Brown, Forman, Orellana, Zhang, Maynor,
  Betts, DeSimone, and Jaeger]{Brown2010a}
E.~Brown, N.~A. Forman, C.~S. Orellana, H.~Zhang, B.~W. Maynor, D.~E. Betts,
  J.~M. DeSimone and H.~M. Jaeger, \emph{Nat. Mat.}, 2010, \textbf{9},
  220--4\relax
\mciteBstWouldAddEndPuncttrue
\mciteSetBstMidEndSepPunct{\mcitedefaultmidpunct}
{\mcitedefaultendpunct}{\mcitedefaultseppunct}\relax
\EndOfBibitem
\bibitem[Brown and Jaeger(2012)]{Brown2012a}
E.~Brown and H.~M. Jaeger, \emph{J. Rheol.}, 2012, \textbf{56}, 875\relax
\mciteBstWouldAddEndPuncttrue
\mciteSetBstMidEndSepPunct{\mcitedefaultmidpunct}
{\mcitedefaultendpunct}{\mcitedefaultseppunct}\relax
\EndOfBibitem
\bibitem[Fall \emph{et~al.}(2012)Fall, Bertrand, Ovarlez, and Bonn]{Fall2012}
A.~Fall, F.~Bertrand, G.~Ovarlez and D.~Bonn, \emph{J. Rheol.}, 2012,
  \textbf{56}, 575\relax
\mciteBstWouldAddEndPuncttrue
\mciteSetBstMidEndSepPunct{\mcitedefaultmidpunct}
{\mcitedefaultendpunct}{\mcitedefaultseppunct}\relax
\EndOfBibitem
\bibitem[Hoffman(1972)]{Hoffman1972}
R.~L. Hoffman, \emph{J. Rheol.}, 1972, \textbf{16}, 155\relax
\mciteBstWouldAddEndPuncttrue
\mciteSetBstMidEndSepPunct{\mcitedefaultmidpunct}
{\mcitedefaultendpunct}{\mcitedefaultseppunct}\relax
\EndOfBibitem
\bibitem[Fernandez \emph{et~al.}(2013)Fernandez, Mani, Rinaldi, Kadau, Mosquet,
  Lombois-Burger, Cayer-Barrioz, Herrmann, Spencer, and Isa]{Fernandez2013}
N.~Fernandez, R.~Mani, D.~Rinaldi, D.~Kadau, M.~Mosquet, H.~Lombois-Burger,
  J.~Cayer-Barrioz, H.~J. Herrmann, N.~D. Spencer and L.~Isa, \emph{Phys. Rev.
  Lett.}, 2013, \textbf{111}, 108301\relax
\mciteBstWouldAddEndPuncttrue
\mciteSetBstMidEndSepPunct{\mcitedefaultmidpunct}
{\mcitedefaultendpunct}{\mcitedefaultseppunct}\relax
\EndOfBibitem
\bibitem[Seto \emph{et~al.}(2013)Seto, Mari, Morris, and Denn]{Seto2013}
R.~Seto, R.~Mari, J.~F. Morris and M.~M. Denn, \emph{Phys. Rev. Lett.}, 2013,
  \textbf{111}, 218301\relax
\mciteBstWouldAddEndPuncttrue
\mciteSetBstMidEndSepPunct{\mcitedefaultmidpunct}
{\mcitedefaultendpunct}{\mcitedefaultseppunct}\relax
\EndOfBibitem
\bibitem[Brown and Jaeger(2014)]{Brown2014}
E.~Brown and H.~M. Jaeger, \emph{Rep. Prog. Phys.}, 2014, \textbf{77},
  046602\relax
\mciteBstWouldAddEndPuncttrue
\mciteSetBstMidEndSepPunct{\mcitedefaultmidpunct}
{\mcitedefaultendpunct}{\mcitedefaultseppunct}\relax
\EndOfBibitem
\bibitem[Liu \emph{et~al.}(2010)Liu, Shelley, and Zhang]{Liu2010}
B.~Liu, M.~Shelley and J.~Zhang, \emph{Phys. Rev. Lett.}, 2010, \textbf{105},
  188301\relax
\mciteBstWouldAddEndPuncttrue
\mciteSetBstMidEndSepPunct{\mcitedefaultmidpunct}
{\mcitedefaultendpunct}{\mcitedefaultseppunct}\relax
\EndOfBibitem
\bibitem[von Kann \emph{et~al.}(2011)von Kann, Snoeijer, Lohse, and van~der
  Meer]{VonKann2011}
S.~von Kann, J.~Snoeijer, D.~Lohse and D.~van~der Meer, \emph{Phys. Rev. E},
  2011, \textbf{84}, 1--4\relax
\mciteBstWouldAddEndPuncttrue
\mciteSetBstMidEndSepPunct{\mcitedefaultmidpunct}
{\mcitedefaultendpunct}{\mcitedefaultseppunct}\relax
\EndOfBibitem
\bibitem[Waitukaitis and Jaeger(2012)]{Waitukaitis2012}
S.~R. Waitukaitis and H.~M. Jaeger, \emph{Nature}, 2012, \textbf{487},
  205--209\relax
\mciteBstWouldAddEndPuncttrue
\mciteSetBstMidEndSepPunct{\mcitedefaultmidpunct}
{\mcitedefaultendpunct}{\mcitedefaultseppunct}\relax
\EndOfBibitem
\bibitem[Roch\'{e} \emph{et~al.}(2013)Roch\'{e}, Myftiu, Johnston, Kim, and
  Stone]{Roche2013}
M.~Roch\'{e}, E.~Myftiu, M.~C. Johnston, P.~Kim and H.~A. Stone, \emph{Phys.
  Rev. Lett.}, 2013, \textbf{110}, 148304\relax
\mciteBstWouldAddEndPuncttrue
\mciteSetBstMidEndSepPunct{\mcitedefaultmidpunct}
{\mcitedefaultendpunct}{\mcitedefaultseppunct}\relax
\EndOfBibitem
\bibitem[Waitukaitis \emph{et~al.}(2013)Waitukaitis, Roth, Vitelli, and
  Jaeger]{Waitukaitis2013}
S.~R. Waitukaitis, L.~K. Roth, V.~Vitelli and H.~M. Jaeger, \emph{Eur. Lett.},
  2013, \textbf{102}, 44001\relax
\mciteBstWouldAddEndPuncttrue
\mciteSetBstMidEndSepPunct{\mcitedefaultmidpunct}
{\mcitedefaultendpunct}{\mcitedefaultseppunct}\relax
\EndOfBibitem
\bibitem[Burton \emph{et~al.}(2013)Burton, Lu, and Nagel]{Burton2013}
J.~C. Burton, P.~Y. Lu and S.~R. Nagel, \emph{Phys. Rev. Lett.}, 2013,
  \textbf{111}, 188001\relax
\mciteBstWouldAddEndPuncttrue
\mciteSetBstMidEndSepPunct{\mcitedefaultmidpunct}
{\mcitedefaultendpunct}{\mcitedefaultseppunct}\relax
\EndOfBibitem
\bibitem[Waitukaitis(2013)]{Waitukaitis}
S.~R. Waitukaitis, \emph{PhD thesis}, University of Chicago, 2013\relax
\mciteBstWouldAddEndPuncttrue
\mciteSetBstMidEndSepPunct{\mcitedefaultmidpunct}
{\mcitedefaultendpunct}{\mcitedefaultseppunct}\relax
\EndOfBibitem
\bibitem[G\'{o}mez \emph{et~al.}(2012)G\'{o}mez, Turner, van Hecke, and
  Vitelli]{Gomez2012}
L.~R. G\'{o}mez, A.~M. Turner, M.~van Hecke and V.~Vitelli, \emph{Phys. Rev.
  Lett.}, 2012, \textbf{108}, 058001\relax
\mciteBstWouldAddEndPuncttrue
\mciteSetBstMidEndSepPunct{\mcitedefaultmidpunct}
{\mcitedefaultendpunct}{\mcitedefaultseppunct}\relax
\EndOfBibitem
\bibitem[G\'{o}mez \emph{et~al.}(2012)G\'{o}mez, Turner, and
  Vitelli]{Gomez2012a}
L.~R. G\'{o}mez, A.~M. Turner and V.~Vitelli, \emph{Phys. Rev. E}, 2012,
  \textbf{86}, 041302\relax
\mciteBstWouldAddEndPuncttrue
\mciteSetBstMidEndSepPunct{\mcitedefaultmidpunct}
{\mcitedefaultendpunct}{\mcitedefaultseppunct}\relax
\EndOfBibitem
\bibitem[Bi \emph{et~al.}(2011)Bi, Zhang, Chakraborty, and Behringer]{Bi2011}
D.~Bi, J.~Zhang, B.~Chakraborty and R.~P. Behringer, \emph{Nature}, 2011,
  \textbf{480}, 355--8\relax
\mciteBstWouldAddEndPuncttrue
\mciteSetBstMidEndSepPunct{\mcitedefaultmidpunct}
{\mcitedefaultendpunct}{\mcitedefaultseppunct}\relax
\EndOfBibitem
\bibitem[Sarkar \emph{et~al.}(2013)Sarkar, Bi, Zhang, Behringer, and
  Chakraborty]{Sarkar2013}
S.~Sarkar, D.~Bi, J.~Zhang, R.~P. Behringer and B.~Chakraborty, \emph{Phys.
  Rev. Lett.}, 2013, \textbf{111}, 068301\relax
\mciteBstWouldAddEndPuncttrue
\mciteSetBstMidEndSepPunct{\mcitedefaultmidpunct}
{\mcitedefaultendpunct}{\mcitedefaultseppunct}\relax
\EndOfBibitem
\end{mcitethebibliography}
}

\end{document}